\documentclass[a4paper,12pt]{article}
\usepackage{graphicx}
\usepackage{amssymb}
\usepackage{amsmath}
\usepackage{cite}

\def\be{\begin{equation}}
\def\ee{\end{equation}}
\def\bea{\begin{eqnarray}}
\def\eea{\end{eqnarray}}
\def\beb{\begin{eqnarray*}}
\def\eeb{\end{eqnarray*}}

\newlength{\myVSpace}
\begin{document}
\makeatletter
\def\fmslash{\@ifnextchar[{\fmsl@sh}{\fmsl@sh[0mu]}}
\def\fmsl@sh[#1]#2{%
  \mathchoice
    {\@fmsl@sh\displaystyle{#1}{#2}}%
    {\@fmsl@sh\textstyle{#1}{#2}}%
    {\@fmsl@sh\scriptstyle{#1}{#2}}%
    {\@fmsl@sh\scriptscriptstyle{#1}{#2}}}
\def\@fmsl@sh#1#2#3{\m@th\ooalign{$\hfil#1\mkern#2/\hfil$\crcr$#1#3$}}
\makeatother
\thispagestyle{empty}
\begin{titlepage}


\boldmath
\begin{center}
{\large {\bf CHARMLESS AND STRANGLESS \\
\vspace {.5cm}
NONLEPTONIC B DECAYS}}
\end{center}
\unboldmath
\vspace{1cm}

\renewcommand{\thefootnote}{\fnsymbol{footnote}}
\begin{center}

{{{\bf
Josip Trampeti\'{c}${}$\footnote{e-mail:
josipt@rex.irb.hr}
}}}
\vspace{1cm}
\end{center}
\setcounter{footnote}{0}
\renewcommand{\thefootnote}{\arabic{footnote}}
\vskip 1em
\begin{center}
Rudjer Bo\v{s}kovi\'{c} Institute, \\
Theoretical Physics Division, \\
        P.O.Box 180, 10002 Zagreb, Croatia\\
 \end{center}

\vspace{2cm}

\begin{abstract}
\noindent
Decay rates of the B-meson are studied through charmless
and strangless transitions into $\pi$, $\rho$ , $\omega$ and $\gamma$ systems.
The important features of these modes are their clean signatures.
The CLEO II collaboration has recently reported the value
$BR(B^{0} \rightarrow \pi^{+} \pi^{-}) = 1.3_{-0.6}^{+0.8}\pm 0.2
\times 10^{-5}$. This value test our approach to nonleptonic B-decays within
the standard model (SM). Since $B \rightarrow \gamma \gamma$
is far beyond our experimental reach, we believe that the
correct determination of the order of magnitude $\sim 10^{-10}$
for $BR(B \rightarrow \gamma \gamma)$ provides the most reliable 
value needed at this moment. The most recent experimental report by the CLEO
collaboration on $BR(B \rightarrow K^{*} \gamma) = 
(4.5 \pm 1.5 \pm 0.9) \times 10^{-5}$
represents a confirmation of our SM prediction
for the $B \rightarrow K^{*} \gamma$ decay. This experimental
result, despite of its great importance for the SM and
the physics beyond the SM, enables us to predict the value 
$BR\left(B \rightarrow \rho\gamma\right)
\cong (3.5 \pm 3.3)\times 10^{-6}$.
\end{abstract}

\end{titlepage}

\section{Introduction}
\label{sec:intro}

Hadronic rare B decays have recently been the subject of both theoretical
and experimental interest. The estimates for these processes involve the 
matrix elements of four quark operators, and these did not receive much
attention in the past. In this paper we propose to study a number of
B decays that follow from $b \rightarrow u,d$ and have clean signatures.
The calculation proceedes in two steps. The effective short-distance 
interaction consists of two contributions. One contribution comes
from the QCD-corrected tree-level w-exchange diagram \cite{Gaillard:1974nj}, with
the $b \rightarrow u$ vertex. The other contribution comes from the
gluon exchanged with $b \rightarrow d$ transitions \cite{Deshpande:1982mi}. The second
step is to use the factorization approximation to derive the hadronic
matrix elements by saturating $H_{w}^{eff}$ with the vacuum
state in all possible ways. The resulting matrix elements involve
quark bilinears between one meson and the vacuum and between
two meson states. These matrix elements are estimated using relativistic
quark model wave functions. Such a technique has been extensively used
for B and D nonleptonic decays by Bauer et al \cite{Wirbel:1985ji} and the results
are consistent with experiment. These methods for $ b \rightarrow s$
nonleptonic modes were employed \cite{Deshpande:1989vu} in our previous study.

The short- and long-distance contribution to $B \rightarrow \rho \gamma$
and the determination of the order of magnitude for
$B \rightarrow \gamma \gamma$ are described in the last two sections 
of this paper.

\section{Effective Hamiltonian}
\label{sec:Eff H}

  The effective weak interaction Hamiltonian (local four-quark operator)
describing the $ b\rightarrow u$ transitions for charmless
and strangeless B-meson decays at the tree level is
\begin{equation}
 H_{eff}^{w} = \sqrt{2}G_{F}\left(c_{-}O_{-} + c_{+}O_{+}\right) + h.c., 
\label{1}
\end{equation}
\begin{equation} 
 O_{\pm} = \overline{u}_{L}^{i}\gamma_{\mu}b_{L}^{i} \overline{d}_{L}^{j}
\gamma^{\mu}u_{L}^{j} {\pm} \overline{d}_{L}^{i}\gamma_{\mu}b_{L}^{i}
\overline{u}_{L}^{j}\gamma^{\mu} u_{L}^{j}.
\label{2}
\end{equation}
Here $c_{\pm}$ are the QCD coefficients and (i, j) are the color indices. The
parameters used to evaluate the $c_{\pm}$ coefficients are
\begin{equation}
 m_{b} = 4.9 GeV,\hspace*{1cm} \Lambda_{QCD} = 200 MeV,
 \hspace*{1cm} \alpha_{s}\left(m_{b}\right) = 0.256,
\label{3}
\end{equation}
\begin{equation}
 c_{-} = 1.34,\hspace*{1cm}  c_{+} = 0.86.  
\label{4}
\end{equation}
  Charmless and strangless B decays can also arise through a one-loop process
involving gluon exchange. The relevant Hamiltonian is \cite{Deshpande:1982mi}
\begin{eqnarray}
 H_{eff}^{P} &=& \kappa \left(\overline{d}_{L}^{i}\gamma_{\mu}b_{L}^{i} 
\overline{q}_{L}^{j}\gamma^{\mu}q_{L}^{j} 
-3 \overline{d}_{L}^{i}\gamma_{\mu}b_{L}^{j} 
\overline{q}_{L}^{j}\gamma^{\mu}q_{L}^{i}
\right.
\nonumber \\
&+&  \left.\overline{d}_{L}^{i}\gamma_{\mu}b_{L}^{i} 
\overline{q}_{R}^{j}\gamma^{\mu}q_{R}^{j} 
-3 \overline{d}_{L}^{i}\gamma_{\mu}b_{L}^{j} 
\overline{q}_{R}^{j}\gamma^{\mu}q_{R}^{i} \right),
\label{5}
\end{eqnarray}
\begin{equation}
 \kappa = \frac{\alpha_{s}G_{F}G_{1}}{6\pi\sqrt{2}}, \hspace*{1cm}
 \alpha_{s} = \frac{g_{s}^{2}}{4\pi}.
\label{6}
\end{equation}
Here q runs over all quark species, although only u and d are relevant
to our discussion. We have used $\alpha_{s}(m_{b}) = 0.256$ as in (\ref{3}), and
$G_{1} = -6.52 V_{bc}V_{uc}^{*}$ corresponding to $m_{t} = 150 GeV$.

\section{Factorization approximation}
\label{sec:Fappr}

  From experience we know that nonleptonic decays are extremely difficult
to handle. For example, the $\Delta I = 1/2$ rule in $K \rightarrow \pi\pi$
decays has not yet been understood in a satisfactory way. Enormous
theoretical machinery has been applied to $K \rightarrow \pi \pi$ decays
producing only up to 50\% agreement with experiment. For energetic decays
of heavy mesons (D, B), the situation is somewhat simpler. For these decays,
the direct generation of a final meson by a quark current is (probably) a good
approximation.

According to the current-field identities, the currents are proportional
to interpolating stable or quasi-stable hadron fields. The approximation
now consists only in taking the asymptotic part of the full hadron
field, i.e. its "in" or "out" field. Then the weak amplitude
factorizes and is fully determined
by the matrix elements of another current between the two remaining
hadron states. For that reason, we call this approximation the
factorization approximation.
  Note that in replacing the interacting fields by the asymptotic
fields, we have neglected any initial or final-state interaction of
the corresponding particles. For B decays, this can be justified by
the simple energy argument that one very heavy object decays into two
light but very energetic objects whose interactions might be safely
neglected. Also, diagrams in which a quark pair is created from the vacuum
will have small amplitudes because  these quarks have to combine with
fast quarks to form the final meson. Note also that the $1/N_{c}$
-expansion argument provides a theoretical justification for the
factorization approximation, since it follows the leading order in
the $1/N_{c}$ expansion. Here $N_{c}$ is the number of colors.

  Each of the B-decay mode might receive three different contributions.
As an example we give one amplitude obtained from $H_{eff}^{w}$ :
\begin{eqnarray}
 A\left(B^{+}(p) \rightarrow \pi^{+}(k) \pi^{0}(q) \right) & =& 
L(\pi^{0}) \langle \pi^{+} \left|\overline{b} \gamma_{\mu}(1-\gamma_{5})d\right|
B^{+}\rangle\langle\pi^{0}\left|\overline{u}\gamma^{\mu}\gamma_{5}u\right|0\rangle
 \nonumber \\ & + & L(\pi^{+})\langle\pi^{+} \left|\overline{u}\gamma_{\mu}\gamma
_{5}d\right|0\rangle\langle\pi^{0}\left|\overline{b}\gamma^{\mu}(1-\gamma_{5})u
\right|B^{+}\rangle \nonumber \\ & + & L(B^{+})\langle\pi^{+}\pi^{0}\left|\overline{u}\gamma_{\mu}
(1-\gamma_{5})d\right|0\rangle\langle0\left|\overline{b}\gamma^{\mu}\gamma_{5}u\right|B^{+}\rangle .
\nonumber\\
\label{7}
\end{eqnarray}
The coefficients $ L(\pi^{0}), L(\pi^{+})$ and $ L(B^{+})$ contain the coupling
constants, color factors, flavor symmetry factors, i.e. flavor counting
factors and factors resulting from the Fierz transformation
of the operators in Eqs.(\ref{1}) and (\ref{4}).
The coefficients $ L(\pi^{0})$ and $ L(\pi^{+})$ correspond to
the quark decay diagram, whereas the $ L(B^{+})$ corresponds to the so-called
annihilation diagrams. These factors are different for each decay mode,
as indicated by the dependence on the final-state meson. To obtain the
amplitudes for other decay modes, one has to replace the final-state
particles with the particles relevant to that particular mode.

The QCD coefficients appear in two different combinations in the amplitudes
of various decay modes \cite{Tadic:1982vn}:
\begin{equation}
C_{1}= \frac{1}{2} \left[c_{+}\left(1+ \frac{1}{N_{c}}\right)
+c_{-}\left(1- \frac{1}{N_{c}}\right)\right],
\label{8}
\end{equation}
\begin{equation}
C_{2}= \frac{1}{2} \left[c_{+}\left(1+ \frac{1}{N_{c}}\right)
-c_{-}\left(1- \frac{1}{N_{c}}\right)\right].
\label{9}
\end{equation}
The factor $1/N_{c}$ arises from the color mismatch in forming color
singlets after Fierz transformation.

We proceed with the definitions of the coupling constants and the
Lorentz decomposition of the typical hadronic matrix elements:
\begin{eqnarray}
\langle 0\left|\overline{b}\gamma_{\mu}\gamma_{5}u\right|B^{+}(p)\rangle = 
ip_{\mu}f_{B}\;,\hspace*{.2cm}
 f_{B} = 1.5f_{\pi}, \hspace*{.2cm} f_{\pi} = 130 MeV,
\label{10}
\end{eqnarray}
\begin{eqnarray}
\langle\rho^{0}(k)\left|\overline{u}\gamma_{\mu}u\right|0\rangle = ig_{\rho^{0}}
\epsilon_{\mu}(k), \hspace*{0.2 cm}
\langle\omega(k)\left|\overline{u}\gamma_{\mu}u\right|0\rangle =
ig_{\omega}\epsilon_{\mu}(k), \hspace*{0.2 cm}
g_{\rho^{0}} = \frac{g_{\rho^{+}}}{\sqrt{2}},
\label{11}
\end {eqnarray}
\begin{eqnarray}
\langle\pi^{+}(k)\left|\overline{b}\gamma_{\mu}d\right|B^{+}(p)\rangle = 
(p+k)_{\mu}f^{(+)}(q^{2}) + q_{\mu}f^{(-)}(q^{2}), 
\hspace*{.2cm} q=p-k
\label{12}
\end{eqnarray}
\begin{eqnarray}
&&\langle\rho^{+}(k)\left|\overline{b}\gamma_{\mu}(1-\gamma_{5})d\right|B^{+}(p)\rangle =
i\varepsilon_{\mu \nu \lambda \sigma}\epsilon^{\nu}(k)(p+k)^{\lambda}
(p-k)^{\sigma}V(q^{2}) \nonumber \\
&+& \varepsilon_{\mu}(k)\left[(m_{B}^{2}-m_{\rho}^{2})A_{1}(q^{2})-
(\varepsilon \cdot q)(p+k)_{\mu}A_{2}(q^{2})\right]\label{13}\\
&+& q_{\mu}(\frac{\varepsilon \cdot q}{q^2})(m_{B}+m_{\rho})
\left[2m_{\rho}A_{0}(q^{2})-(m_{B}-m_{\rho})\left(A_{1}(q^{2})-A_{2}(q^{2})
\right)\right],
\nonumber 
\end{eqnarray}
\begin{equation}
\langle\pi^{+}(k)\rho^{0}(q)\left|\overline{u}\gamma_{\mu}d\right|0\rangle =
\langle\rho^{0}(q)\left|\overline{u}\gamma_{\mu}d\right|\pi^{-}(-k)\rangle.
\label{14}
\end{equation}

Any hadronic matrix element needed to evaluate the branching ratios of the
decays can easily be obtained from the above definitions.
We assume that the momentum dependence of the form-factors $f^{+}(q^{2})$,
$V(q^{2})$, etc. from Eqs.(\ref{11}-\ref{13}), is well described by single poles with
masses of excited b-quark meson states $(1^{-}, 0^{-}, ...)$ close to $m_{B}$,
i.e. we can use only the nearest pole dominance.
The form factors at zero momentum transfer were calculated using the
relativistic oscillator wave functions \cite{Wirbel:1985ji}. Now we give a few examples
of the structure of the form-factors:
\begin{eqnarray}
f_{\pi^{+}B}^{(+)}\left(m_{\pi}^{2}\right) & =& \frac{f_{\pi^{+}B}^{(+)}(0)}
{\left(1-\frac{m_{\pi}^{2}}{m_{0^{+}}^{2}}\right)}\;, 
\label{15}
\end{eqnarray}
\begin{eqnarray}
V_{\rho^{+}B}\left(m_{\rho}^{2}\right) & = & \frac{V_{\rho^{+}B}(0)}
{(m_{B}+m_{\rho})\left(1-\frac{m_{\rho}^{2}}{m_{1^{-}}^{2}}\right)}\;, 
\label{16}
\end{eqnarray}
\begin{eqnarray}
A_{0}^{\rho^{+}B}\left(m_{\rho}^{2}\right) & = & \frac{A_{0}^{\rho^{+}B}(0)}
{(m_{B}+m_{\rho})\left(1-\frac{m_{\rho}^{2}}{m_{0^{-}}^{2}}\right)}\;, 
\label{17}
\end{eqnarray}
\begin{eqnarray}
A_{1}^{\rho^{+}B}\left(m_{\rho}^{2}\right) & = & \frac{A_{1}^{\rho^{+}B}(0)}
{(m_{B}-m_{\rho})\left(1-\frac{m_{\rho}^{2}}{m_{1^{+}}^{2}}\right)}\;, 
\label{18}
\end{eqnarray}
\begin{eqnarray}
A_{2}^{\rho^{+}B}\left(m_{\rho}^{2}\right) & = & \frac{A_{2}^{\rho^{+}B}(0)}
{(m_{B}+m_{\rho})\left(1-\frac{m_{\rho}^{2}}{m_{1^{+}}^{2}}\right)}\;. 
\label{19}
\end{eqnarray}
Here we have used
$m_{0^{-}}(b) \approx m_{0^{+}}(b) \approx m_{1^{-}}(b) \approx
m_{1^{+}}(b) \approx m_{B}$. The couplings $g_{\rho}$, $g_{\omega}$ are 
determined from the $\rho, \omega \rightarrow e^{+}e^{-}$ experimental rates:
\begin{eqnarray}
g_{\rho^{0}}^{2} & = & \frac{3m_{\rho}^{3}}{4\pi\alpha^{2}}
\Gamma(\rho \rightarrow e^{+}e^{-}) = 0.0141 GeV^{4}\;, 
\label{20}
\end{eqnarray}
\begin{eqnarray}
g_{\omega}^{2} & = & \frac{27m_{\omega}^{3}}{4\pi\alpha^{2}}
\Gamma(\omega \rightarrow e^{+}e{-}) = 0.0128 GeV^{4}.
\label{21}
\end{eqnarray}

For the ratio of the Kobayashi-Maskawa (K-M) matrix elements
we use the central value
$\left|V_{ub}/V_{bc}\right| = 0.1\zeta$, where $0<\zeta<1$.
Some examples of the branching ratios that arise from the tree-level
Hamiltonian are
\begin{equation}
BR\left(B^{0} \rightarrow \pi^{+}\pi^{-}\right)  =  2\left(\frac
{\pi f_{\pi}}{m_{B}}\right)^{2}\left|\frac{V_{ub}}{V_{bc}}
\right|^{2}\left|C_{1}\right|^{2}\left|f_{B
\pi}^{(+)}\left(m_{\pi}^{2}\right)\right|^{2}\lambda_{\pi\pi}^{1/2}, 
\label{22}
\end{equation}
\begin{equation} 
BR\left(B^{0} \rightarrow \rho^{0}\omega\right)  =  
\left(\frac{\pi g_{\rho^{0}}}{m_{B}}\right)^{2}\left|
\frac{V_{ub}}{V_{bc}}\right|^{2} 
\left|C_{2}\right|^{2}\left(1+\frac{g_{\omega}}{g_{\rho^{0}}}\right)^{2}
\left|F\right|\lambda_{\rho\omega}^{3/2},  
\label{23}
\end{equation}
\begin{equation} 
 BR\left(B^{+} \rightarrow \rho^{+}\omega\right)  =
\left(\frac{\pi g_{\rho^{+}}}{m_{B}}\right)^{2}\left|
\frac{V_{ub}}{V_{bc}}\right|^{2}
\left|C_{1}-\frac{g_{\omega}}{g_{\rho^{0}}}C_{2}\right|^{2}
\left|F\right|\lambda_{\rho\omega}^{3/2},
\label{24}
\end{equation}
where
\begin{eqnarray}
\lambda_{ab} &= &\left(1-\frac{m_{a}^{2}}{m_{B}^{2}}-\frac{m_{b}^{2}}{m_{B}^{2}}\right)^{2}
-\frac{4m_{a}^{2}m_{b}^{2}}{m_{B}^{4}}\;,
\label{25}
\end{eqnarray}
\begin{eqnarray}
\left|F\right| & =& 2V^{2}\left(m_{\rho}^{2}\right) 
\nonumber \\
&-& \frac{m_{B}^{4}}{2m_{\rho}^{4}}\left(1-\frac{2m_{\rho}^{2}}{m_{B}^{2}}\right)
\left(1-\frac{m_{\rho}^{2}}{m_{B}^{2}}\right)A_{1}\left(m_{\rho}^{2}\right)
A_{2}\left(m_{\rho}^{2}\right)
\nonumber \\
&+& \left(\frac{3}{\lambda_{\rho\rho}} 
+ \frac{m_{B}^{4}}{4m_{\rho}^{4}}\right)\left(1-\frac{m_{\rho}^{2}}{m_{B}^{2}}\right)^{2}
A_{1}^{2}\left(m_{\rho}^{2}\right) 
  +  \frac{m_{B}^{4}}{4m_{\rho}^{4}}A_{2}^{2}\left(m_{\rho}^{2}\right) \;.
\label{26}
\end{eqnarray}

  The results for all the modes using the tree-level Hamiltonian are given in
Table 1. The values of the branching ratios presented
in Table 1 have been obtained by applying the
$N_{c} \rightarrow \infty$ limit, i.e. for $C_{1} = 1.1$
and $C_{2} = -0.24$.  We find that the penguin contributions to these
decays are generally small. We therefore estimate the contribution
to a few of the more interesting modes given in Table 1. Finally note that the
branching ratios BR are evaluated by normalizing the partial
widths $\Gamma$ to the B-meson lifetime, 
i.e. to the total B-meson decay width 
(for details, see Refs.\cite{Deshpande:1988pi,RPP}):
\begin{equation}
\Gamma_{tot}(B) = \frac{1}{\tau_{B}} \cong 
3 \frac{G_{F}^{2} m_{B}^{5}\left|V_{cb}\right|^{2}}{192 \pi^{3}} .
\label{27}
\end{equation}

\newpage
\centerline{\it TABLE 1 }
\vspace{.3cm}
The branching ratios of various charmless 
and strangless nonleptonic B-decay modes. 
\vspace{.5cm}\\
\begin{tabular}{|c|c|c|c|c|c|} \hline \hline
Branching& &QCD coeff.&Contribution&Contribution&ARGUS/CLEO \\
Ratio&Ref.[3]&[for the tree&from $H_{eff}^{w}$&from $H_{eff}^{P}$&
Exp. limit [6] at \\
Mode &$[10^{-5}]\zeta^{2}$&level graphs]&$[10^{-5}]\zeta^{2}$& $[10^{-5}]$
& 90 \% CL $[10^{-5}]$ \\ \hline \hline
$B^{+} \rightarrow \pi^{+}\pi^{0}$ & 0.6 & $C_{1}-C_{2}$& 0.49& 0.00& 50/260 \\
$\pi^{+}\rho^{0}$& 0.6&$C_{2}$&0.18& & 19/17 \\
$\pi^{+}\omega$& &$C_{2}$&0.18& & \\
$\pi^{0}\rho^{+}$&0.6&$C_{1}$&1.74& & \\ 
$\rho^{+}\rho^{0}$&1.4& $C_{1}-C_{2}$&1.15& & \\
$\rho^{+}\omega$& &Eq.(24)&1.19& & \\
\hline
$B^{0}\rightarrow\pi^{+}\pi^{-}$ & 2.1 & $C_{1}$ & 1.56 & $3\times 10^{-4}$&
$1.3_{-0.6}^{+0.8} \pm 0.2$ [7] \\
$\pi^{0}\pi^{0}$ & & $C_{2}$ & 0.03 & $2.5\times 10^{-4}$ & \\ 
$\pi^{+}\rho^{-}$ & 5.6 & $C_{1}$ & 4.35 & & \\
$\pi^{-}\rho^{+}$ & & $C_{1}$ & 1.07 & & \\
$\pi^{0}\rho^{0}$ & 0.1 & $C_{2}$ & 0.10 & $1.8\times 10^{-3}$ & 43 \\
$\pi^{0}\omega$ & & $C_{2}$ & 0.09 & & \\
$\rho^{+}\rho^{-}$ & 4.5 & $C_{1}$ & 3.60 & $4\times 10^{-3}$ & 420 \\
$\rho^{0}\rho^{0}$ & 0.1 & $C_{2}$ & 0.07 & & \\
$\rho^{0}\omega$ & & $C_{2}$ & 0.14 & & \\
$\omega\omega$ & & $C_{2}$ & 0.06 & & \\
\hline \hline
\end{tabular}

\section{Short and long distance contributions to $B\rightarrow\rho\gamma$} 
\label{sec:BRG}

In April this year the CLEO collaboration announced the
experimental discovery of electromagnetic-penguin B decays
$B\rightarrow K^{*}\gamma$. The information from
Wilson Lab on 13-APR-1993 was the following \cite{Stone} :

Using a data sample of $2.8\times10^{6}$ B-meson decays collected by the CLEO II
detector operating at the Cornell Electron Storage Ring (CESR) it was possible 
to observe the rare b-quark decay $b\rightarrow s\gamma$ in the modes
$B^{0} \rightarrow {K^{*}}^{0}(892) \gamma$ and $B^{-} \rightarrow {K^{*}}^{-}(892)
\gamma$ (the charge conjugation was implied throughout).
Details are given below in Table 2. This is the first unambiguous
evidence for penguin-type diagrams in weak decays.
\newpage
\centerline{\it TABLE 2}
\vspace{.3cm} 
The CLEO collaboration experimental discovery of 
electromagnetic-\\penguin B-decays [10]: $B \rightarrow K^{*} \gamma$. 
\vspace{.5cm}\\
\begin{tabular}{|c||c|c|c|} \hline \hline
Mode &{${K^*}^{0} \gamma$} &
  \multicolumn{2}{|c|}{${K^*}^{-} \gamma$} \\   
\hline \hline
$K*$ decay mode & $K^{-}\pi^{+}$ & $K_{s}^{0}\pi^{-}$ & $K^{-}\pi^{0}$ \\
\hline
Signal events & 8 & 2 & 3 \\
Sideband events & 41 & 2 & 10 \\
Sideband scale factor$^{(*)}$ & 1.0/37.6 & 1.0/40. & 1.0/12. \\
\hline
Extrapolated background & $1.1\pm0.2$ & $0.050\pm0.035$ & $0.83\pm0.26$ \\
Binomial probability of & & & \\
background fluctuation & $3.7\times 10^{-5}$ & 0.35\% & 7.30\% \\
\hline
Efficiency & $(11.9\pm1.8)$ \% & 2.0\% & 3.1\% \\
Additional $B\overline{B}$ backgnd$^{(**)}$ & 0.30$\pm$0.15 & 0.01 & 0.10 \\
 \hline \hline
Branching ratio [$10^{-5}$] & 4.0 $\pm$1.7 $\pm$ 0.8 & \multicolumn{2}{|c|}{5.7 $\pm$ 3.1 $\pm$ 1.1} \\
\hline \hline
\multicolumn{2}{|c|}{Combined BR $[B \rightarrow ( {K^*}^{0} + {K^*}^{-})
 \gamma]$} &  \multicolumn{2}{|c|}{$(4.5 \pm 1.5 \pm 0.9) \times 10^{-5}$} \\
\hline \hline 
\multicolumn{4}{l}{
(*)This is the number of background events in the signal region 
} \\
\multicolumn{4}{l}{
divided by the number of  background events in the sideband 
} \\
\multicolumn{4}{l}{
region from off $B\overline{B}$ resonance data. 
} \\
\multicolumn{4}{l}{
(**) This is the number of background events expected from $B\overline{B}$ 
} \\
\multicolumn{4}{l}{
events from the Monte Carlo  simulation. Note that the background 
} \\
\multicolumn{4}{l}{
is dominated by non-$B\overline{B}$ processes. 
} \\
\end{tabular} \\

The branching fractions were calculated by summing signal events, subtracting
summed background events and dividing by summed efficiencies. In particular,
the combined ${K^*}^{0}$ and ${K^*}^{-}$ number was NOT obtained by averaging
the individual results, weighted by the statistical (or any other) error. The
combined number was obtained by summing all signal events, subtracting all
backgrounds and dividing by the sum of all efficiencies.

The short distance (SD) contribution to $ B \rightarrow K^{*} \gamma$
was proposed and obtained in several papers 
\cite{Deshpande:1987nr,Grinstein:1990tj}.
The branching ratio obtained is of the form
\begin{equation}
BR^{SD}(B\rightarrow K^{*} \gamma) = \frac{\alpha}{2\pi}
\left(\frac{m_{b}}{m_{B}}\right)^{2} \left|\frac{V_{ts}^{*}V_{tb}}
{V_{bc}}\right|^{2} \left|f_{1}^{BK^{*}}(0)\right|^{2}
\left|\tilde{F}_{2}^{bs}(m_{t})\right|^{2}.
\label{28}
\end{equation}
Here $\tilde{F}_{2}^{bs}(m_{t})$ is the one-loop flavor-changing
$b \rightarrow s \gamma$ vertex with QCD corrections,
and $f_{1}^{BK^{*}}(0)$ is the form-factor of the
operator $ \bar{s} \sigma_{\mu \nu} q^{\nu} b_{R}$ between the spin-one
kaon resonance and the B meson, responsible  for the $b \rightarrow s \gamma$
transition at $q^{2} = o$, i.e. for the real photon. Details of the calculations
are given in Ref. \cite{Deshpande:1988pi}. The $BR^{SD}$ is obtained by normalizing the
partial width $\Gamma^{SD} (B \rightarrow K^{*}
\gamma)$ to the B-meson lifetime (see Eq. (\ref{28})). For the top-quark mass
$m_{t} = 150$ GeV, we have found \cite{Deshpande:1988pi}:
\begin{equation}
BR^{SD}(B\rightarrow K^{*} \gamma) = 2.6\times 10^{-5}, 
\label{29}
\end{equation}
which is in excellent agreement with the recent theoretical estimate \cite{Casalbuoni:1993nh}
and is within the error of the measured branching ratio published
recently by the CLEO collaboration \cite{Stone}. 

The short distance contribution to $B \rightarrow \rho \gamma$ is dominated
by the flavor-changing vertex $b \rightarrow d \gamma$ , which proceeds in 
one loop through the exchange of u,c and t quarks and the W-boson. For the
emission of the real photon the only contributing operator 
$ \bar{d} \sigma_{\mu \nu} q^{\nu} b_{R}$ gives the following branching
ratio:
\begin{equation}
BR^{SD}(B\rightarrow \rho \gamma) = \frac{\alpha}{2\pi}
\left(\frac{m_{b}}{m_{B}}\right)^{2} \left|\frac{V_{td}^{*}V_{tb}}
{V_{bc}}\right|^{2} \left|f_{1}^{B\rho}(0)\right|^{2}
\left|\tilde{F}_{2}^{bd}(m_{t})\right|^{2}.
\label{30}
\end{equation}
Here $\tilde{F}_{2}^{bd}(m_{t})$ is the one-loop flavor-changing
$b \rightarrow d \gamma$ vertex with QCD corrections, and the
$f_{1}^{B\rho}(0)$ is the form-factor of the operator
$ \bar{d} \sigma_{\mu \nu} q^{\nu} b_{R}$ between the 
$\rho$-meson and the B-meson. The $BR^{SD}$ is
obtained by normalizing the partial width
$\Gamma^{SD} (B \rightarrow \rho \gamma)$ to the B-meson lifetime.

Careful analysis of the procedure for the one-loop QCD calculations of
the $\tilde{F}_{2}^{bs}(m_{t})$ coefficient 
\cite{Deshpande:1987nr,Grinstein:1990tj} shows that up to
the mass symmetry-breaking effects, one can write
\begin{equation}
\tilde{F}_{2}^{bs}(m_{t}) = \tilde{F}_{2}^{bd}(m_{t}).
\label{31}
\end{equation} 

The calculations of the matrix element
$\langle{K^*}^{+}(k)\left|\overline{s}\sigma_{\mu\nu}q^{\nu}b_{R}\right|B^{+}(p)\rangle$, 
in the Relativistic Constituent Quark Model (RCQM) \cite{Deshpande:1988pi}
gives approximately the same value as the calculation of this matrix
element \cite{Casalbuoni:1993nh} in an effective chiral theory for mesons
using flavor and spin symmetries of the Heavy Quark
Effective Theory [HQET]. Detailed analysis of both calculations
shows that up to the quark mass difference $m_{s}-m_{d}$,
(which is negligible compared to the b-quark mass), we have
\begin{equation}
\langle{K^*}^{+}(k)\left|\overline{s}\sigma_{\mu\nu}q^{\nu}b_{R}\right|B^{+}(p)\rangle =
\langle\rho^{+}(k)\left|\overline{d}\sigma_{\mu\nu}q^{\nu}b_{R}\right|B^{+}(p)\rangle.  
\label{32}
\end{equation} 
Equation (\ref{32}) in the RCQM then gives 
\begin{equation}
f_{1}^{BK^{*}}(0) = f_{1}^{B\rho}(0), \hspace*{1cm}
f_{2}^{BK^{*}}(0) = f_{2}^{B\rho}(0), \hspace*{1cm}
f_{2}^{BK^{*}}(q^{2}) = \frac{1}{2}f_{1}^{BK^{*}}(q^{2}).
\label{33}
\end{equation}
From Eqs. (\ref{28}), (\ref{30}), (\ref{31}) and (\ref{33}) it is easy to find that
\begin{equation}
BR^{SD}(B \rightarrow \rho \gamma) \cong \left| \frac{V_{td}}{V_{ts}}
\right|^{2} BR^{SD}(B \rightarrow K^{*} \gamma).
\label{34}
\end{equation}
The ratio of the K-M matrix elements \cite{RPP} 
$\left|V_{td}/V_{ts}\right|$ = 0.10 to 0.33 and the last line in 
Table 2 gives the following range of values
for $BR^{SD}(B \rightarrow \rho \gamma)$:
\begin{equation}
BR^{SD}(B \rightarrow \rho \gamma) \cong (4.5\;{\rm to}\;50) \times 10^{-7}. 
\label{35}
\end{equation}

It is appropriate here to comment on the long-distance (LD) contribution
to $B \rightarrow \rho \gamma$. The long-distance contribution to
$B \rightarrow K^{*} \gamma$ is discussed in detail in \cite{Deshpande:1988ba}. 
Despite of the conclusion in Ref.\cite{Deshpande:1988ba} that the rate for
$B \rightarrow K^{*} \psi$ cannot be used to give a unique value for
$B \rightarrow K^{*} \gamma$, it is necessary to  evaluate $BR^{LD}
(B \rightarrow \rho \gamma)$ by applying exactly the same procedure,
as in Ref.\cite{Deshpande:1988ba}, as a check of consistency. We use the vector-meson
dominance which leads to the vector-meson - photon conversion mechanism.
This mechanism is known as a long-distance effect.

Owing to the gauge-condition and current-conservation requirements,
for the real outgoing photon in the $B \rightarrow \rho \gamma$ mode we
have that \cite{Deshpande:1988pi} $A_{0} = 0$ and $A_{2} = A_{1} \equiv A$. This gives  
\begin{eqnarray}
BR^{LD}\left(B^{0} \rightarrow \rho^{0}\gamma\right) &=& \left(
\frac{\pi e g_{\rho^{0}}^{2}}{m_{\rho}^{2}m_{B}^{2}}\right)^{2}\left|
\frac{V_{ub}}{V_{bc}}\right|^{2}
\left(1-\frac{m_{\rho}^{2}}{m_{B}^{2}}\right)^{3}
\left(V^{2}(0)+A^{2}(0)\right)
\nonumber\\
&\times&
\left|C_{2}\right|^{2} \left(1+ \frac{g_{\omega}^{2}
m_{\rho}^{2}}{g_{\rho^{0}}^{2}m_{\omega}^{2}}\right)^{2} ,
\label{36}
\end{eqnarray}
\begin{eqnarray}
 BR^{LD}\left(B^{+} \rightarrow \rho^{+}\gamma\right)& =& \left(
\frac{\pi e g_{\rho^{+}}^{2}}{m_{\rho}^{2}m_{B}^{2}}\right)^{2}
\left|\frac{V_{ub}}{V_{bc}}\right|^{2}\left(1-\frac{m_{\rho}^{2}}{m_{B}^{2}}\right)^{3} 
\left(V^{2}(0)+ A^{2}(0)\right)
\nonumber \\
& \times & \left|C_{1}\left(1+\frac{g_{\omega}m_{\rho}^{2}}{g_{\rho^{0}}
m_{\omega}^{2}}\right) - C_{2}\left(1+\frac{g_{\omega}^{2}m_{\rho}^{2}}
{g_{\rho^{0}}^{2}m_{\omega}^{2}}\right)\right|^{2}\;.
\label{37}  
\end{eqnarray}

In the $N_{c} \rightarrow \infty$ limit, i.e. for $C_{1} =1.1$ and
$C_{2} =-0.24$ and for $\left|V_{ub}/V_{bc}\right|=0.1$, we have
\begin{equation}
BR^{LD}\left(B^{0} \rightarrow \rho^{0}\gamma\right) =
BR^{LD}\left(B^{0} \rightarrow \omega\gamma\right) \cong 0.02 \times 10^{-8},
\label{38}
\end{equation}
\begin{equation}
BR^{LD}\left(B^{+} \rightarrow \rho^{+}\gamma\right) \cong 1.4 \times 10^{-8}.
\label{39}
\end{equation}
Assuming that the recently quoted result \cite{Cassel} $\left|V_{ub}/V_{bc}\right|\cong
0.08$ is of reasonable accuracy, the above
$BR^{LD}(B \rightarrow \rho \gamma)$ should become smaller by at least
a factor of two. This smaller branching ratio, together with Eq.(\ref{35}), gives
\begin{equation}
BR^{SD}(B \rightarrow \rho \gamma) \stackrel{>}{\sim} 60\times BR^{LD}(B \rightarrow \rho \gamma).
\label{40}
\end{equation}
This is similar to our previous result \cite{Deshpande:1988ba} obtained with the top quark
mass $m_{t} = 150 GeV$ for $B \rightarrow K^{*}\gamma$:
\begin{equation}
BR^{SD}(B \rightarrow K^{*}\gamma) \stackrel{>}{\sim} 50\times BR^{LD}(B \rightarrow K^{*}\gamma),
\label{41}
\end{equation}
In the best case, the LD contributions to the total branching  ratios
represent $\stackrel{<}{\sim} 2$ \% corrections for both 
$B \rightarrow K^{*} \gamma$ and $B \rightarrow \rho \gamma$.

The similarity of the results in Eqs. (\ref{40}) and (\ref{41}) shows that the 
procedure used in this paper to obtain
$BR^{LD}(B \rightarrow \rho \gamma)$
is consistent with the procedure in Ref.\cite{Deshpande:1988ba}. This means 
that all conclusions drawn in \cite{Deshpande:1988ba} are directly
applicable to this work and that the rates for 
$B \rightarrow \rho \rho, \rho \omega, \omega \omega$ cannot be used
to give a unique value  for $B \rightarrow \rho \gamma$, i.e. Eq.(\ref{35})
fully represents our prediction for the decay rate:
\begin{equation}
2 \times 10^{-7} < BR(B \rightarrow \rho \gamma) < 7 \times 10^{-6}. 
\label{42}
\end{equation}
Here we have also take into account the errors in the measurements \cite{Stone}
of $B \rightarrow K^{*} \gamma$ from Table 2.

\section{$B^{0}$ decay into the two-photon final state}
\label{sec:BGG}

In this case, we should only estimate the order of magnitude for 
$BR(B \rightarrow \gamma \gamma)$.  

First, applying the vector-meson - photon conversion mechanism to the 
neutral B decay modes $(\rho^{0} \rho^{0}, \rho^{0} \omega, \omega \omega)$,
we may evaluate $BR^{LD} (B \rightarrow \gamma \gamma)$ as\\
\begin{equation}
BR^{LD}\left(B^{0} \rightarrow \gamma\gamma\right) =
\left(\frac{\pi e^2 g_{\rho^{+}}^{3}}{m_{B}^{2}m_{\rho}^{4}}\right)^{2}
\left| \frac{V_{ub}}{V_{bc}}\right|^{2}\left|C_{2}\right|^{2}
\left(1+\frac{g_{\omega}^{3}
m_{\rho}^{4}}{g_{\rho^{0}}^{3}m_{\omega}^{4}}\right)^{2}
\left(V^{2}(0)+A^{2}(0)\right). 
\label{43}
\end{equation}
For $C_{2} =-0.24$ (in the $N_{c} \rightarrow \infty$ limit)
and $\left|V_{ub}/V_{bc}\right|=0.1$, we have
\begin{equation}
BR^{LD}\left(B^{0} \rightarrow \gamma\gamma\right) \cong 10^{-12}. 
\label{44}
\end{equation} 

Next, comparing the experimentally measured ratio of widths for
$K_{s}$ decays \cite{RPP} with the ratio of our theoretical estimates from
Eqs.(\ref{38},\ref{44}) and Table 1, we find that 
\begin{equation}
\begin{array}{clcr}
[\Gamma ( K_{s} \rightarrow \gamma \gamma) / \Gamma ( K_{s} \rightarrow
 \pi^{+} \pi^{-})]_{exp} \cong 3.5 \times 10^{-6},
\end{array}
\label{45}
\end{equation}
\begin{equation}
\begin{array}{clcr} 
[\Gamma ( B^{0}\rightarrow \gamma \gamma) / \Gamma ( B^{0} \rightarrow
 \pi^{+} \pi^{-})]_{th} \cong 6.4 \times 10^{-6}. 
\end{array}
\label{46}
\end{equation}
Since the two rates are in reasonable agreement, it is justified to use 
the conversion mechanism to estimate the order of magnitude
for the $B^{0}\rightarrow \gamma \gamma$ rate to be
\begin{equation}
BR\left(B^{0} \rightarrow \gamma\gamma\right) \sim 10^{-10}. 
\label{47}
\end{equation}

\section{Discussion and conclusion}
\label{sec:DC}

One can see from Table 1 that our values are considerably lower than the 
present limits obtained at $90$ \% CL. On the other hand, our results are
in reasonable agreement with the predictions of Ref. 3 summarized in
the second column in Table 1. Note also that the rates involving the
(helicity-suppressed) combination $(c_{-} - 2c_{+})^ 2 $
(or  $(c_{-} - c_{+})^ 2$ in the $N_{c} \rightarrow \infty $ limit)
are sensitive to the values of the QCD coefficients.
Some experimental limits \cite{Avery:1989qi,Kim} are also presented in Table 1.

The CLEO II (Ref.\cite{Kim}) collaboration has recently reported the
$BR(B^{0} \rightarrow \pi^{+} \pi^{-}) < 3.0 \times 10^{-5}$
at $90$ \% CL. This is a significant improvement upon the old results (CLEO 1.5)
\\$BR(B^{0} \rightarrow \pi^{+} \pi^{-}) < 9.0 \times 10^{-5}$ and is welcome
because it is a test of our approach within the standard model.

Before discussing $B \rightarrow \rho \gamma$, let as briefly comment
on $B \rightarrow \gamma \gamma$.
Since the $B \rightarrow \gamma \gamma$ decay is far beyond our
experimental reach, we believe that the correct determination of
the order of magnitude $\sim 10^{-10}$ for $BR(B \rightarrow \gamma \gamma)$
provides the most suitable value needed at this moment.

The most recent CLEO report \cite{Stone} on $BR(B \rightarrow K^{*} \gamma) =
(4.5 \pm 1.5 \pm 0.9) \times 10^{-5}$ represents a confirmation of our
SM prediction for the $B \rightarrow K^{*} \gamma$ decay given six years
ago \cite{Deshpande:1988pi}. This experimental result, despite of its great importance for
the SM and the physics beyond the SM,
enables us to predict the following value for the
$B \rightarrow \rho\gamma$ decay within the SM:
\begin{equation}
BR\left(B \rightarrow \rho\gamma\right) \cong 
(3.5 \pm 3.3) \times 10^{-6}.
\label{48}
\end{equation}
The range of the order of magnitude in Eqs.(\ref{42}) and/or (\ref{48}) is due to the
range of values in the $\left|V_{td}/V_{ts}\right|$ ratio \cite{RPP}.

Assuming that some other experiment might determine the ratio
$\left|V_{td}/V_{ts}\right| \cong 0.1$, 
then at least $10^{8}$ $B^{0}\bar{B}^{0}$ pairs have to be produced at
CESR and collected by the CLEO detector to discover the
electromagnetic penguin B-decays $B \rightarrow \rho \gamma$. It is clear
that $10^{8}$ B-meson decays cannot be collected in the near future.
However, it is encouraging that Cornell has an asymmetric B Factory \cite{CERN6}
proposed as a further CESR upgrade.

On the other hand, if $\left|V_{td}/V_{ts}\right|$ turns out to be
$\cong 0.3$, then only $10^{7}$ $B^{0}\bar{B}^{0}$ pairs have to be
collected by the CLEO detector to discover the $B\rightarrow \rho \gamma$ decay.
This scenario is much more promising because Cornell is currently 
upgrading the luminosity of CESR by at least a factor of five \cite{CERN10}.

In the case of known top-quark mass, the measurement of 
$BR(B \rightarrow \rho \gamma)$ is an excellent 
tool for fixing the weak flavor-mixing parameters more precisely, such as
the $\left|V_{td}/V_{ts}\right|$ ratio.

As is well known, radiative B-meson decays are of importance as a test
of the SM electroweak theory in one loop, because they verify the gauge
structure of the theory. In addition, these decays might contribute
to the discovery of new physics beyond the SM. Finally, note that the
relative smallness of the branching ratio for $B\rightarrow \rho \gamma$
makes it more suitable for possible discovery of new physics than
might be expected from the recently measured branching ratio for
$B \rightarrow K^{*} \gamma$.

\end{document}